\documentclass[10pt]{article}
\usepackage[utf8]{inputenc}
\usepackage{float}
\usepackage{geometry}
\usepackage{hyperref}
    \hypersetup{
    colorlinks=false,
    linkcolor=darkgray,
    filecolor=magenta,      
    urlcolor=cyan,
    }
\usepackage[backend=biber,style=numeric,sorting=none]{biblatex}
    \usepackage{tabularx}
    \usepackage[table]{xcolor}
    \usepackage{multirow}

\usepackage{graphicx}
\graphicspath{ {./images/} }

\usepackage{comment}

\begin{document}

\title{A new Privacy Preserving and Scalable Revocation Method for Self Sovereign Identity - The Perfect Revocation Method does not exist yet}

\author{Andreas Freitag, University of Vienna}
\date{DRAFT November 2022}

\maketitle

\noindent\textbf{Abstract:}\\

Digital Identities are playing an essential role in our digital lives. Today, used Digital Identities are based on central architectures. Central Digital Identity providers control and know our data and, thereby, our Identity. Self Sovereign Identities (SSI) are based on a decentralized data storage and data exchange architecture, where the user is in sole control of his data and identity. Most of the issued credentials need the possibility of revocation. For a Central Digital Identity, revocation is easy. In decentral architectures, revocation is more challenging. Revocation can be done with different methods e.g. lists, compressed lists and cryptographic accumulators. A revocation method must be privacy preserving and must scale. This paper gives an overview about the available revocation methods, include a survey to define requirements, assess different revocation groups against the requirements, highlights shortcomings of the methods and introduce a new revocation method called Linked Validity Verifiable Credentials.

\section{Introduction}
This paper make an assessment of existing revocation methods for Verifiable Credentials (VC) and a new suggested revocation method, named Linked Validity Verifiable Credential (LVVC). The assessment covers privacy, scalability and maturity aspects for the use in Self Sovereign Identity (SSI) systems.  Existing revocation methods are organized in groups. The concept of LVVC is described. In a survey, opinions on privacy and technical requirements for SSI have been collected within the SSI community. From the survey, requirements are derived for the assessment. Based on the requirements, an assessment of the different revocation groups has been done.

\subsection{Self Sovereign Identity}
Digital Identities (DI) are playing an increasingly important role.  In the 1960s, together with databases, DI emerged. In this day's DI have been simple access lists. In the late 1990s Microsoft introduced the first shared DI called Microsoft Passport. It was the predecessor of today Microsoft account. The goal was to have one single identity to access different websites for e-commerce. All common DI have a significant disadvantage compared with physical identity documents. DI systems rely on central entities that control the system and, therefore, the identity of the user. This leads to privacy, controll and access issues. The shortcomings of central DI lead to the concept of SSI. The term SSI was minted and explained first by Christopher Allen in 2016 \cite{allen_path_2016} together with a history of DI and his 10 principles of SSI. Documents that prove identity (e.g. passport) or belong to an identity (e.g. university certificates) and can be verified from a verifier are called VC.
One important SSI principle is the sole control over the identity from the Holder \cite{allen_path_2016} \cite{cameron_laws_2005}. This includes all issued credentials. To be in control, the Holder needs to store credentials on his device. 

\subsection{Roles in an SSI system}
There are three main roles in an SSI system. The Issuer, the Holder and the Verifier \cite{w3c_verifiable_nodate}.
The Issuer is issuing a VC and can perform a revocation of a VC. The Holder is in possession of a VC and can present a VC and a proof of non-revocation to a Verifier. The Verifier asks for information and verifies it. The roles and the interaction between them is called "the triangle of trust". In any DI system, a trust layer is needed. A trust layer can be a Public Key Infrastructure (PKI) or a Decentralized Public Key Infrastructure (DPKI). A PKI or DPKI provides data which is needed for verification, e.g. the public key from the Issuer.

\subsection{Revocation of an SSI Verifiable Credential}
For many VC, the possibility of revocation is essential. The most common example is the driving licence. The simplest way to implement revocation are central allow- or blocklists maintained by the Issuer. But this method contradicts the principles of SSI. The Verifier contacts the Issuer each time he verifies a credential. The Issuer knows when the credential is used and also knows the Verifier. Privacy preserving and scalable revocation in an SSI solution must be solved, otherwise SSI systems cannot be implemented or compromised in terms of privacy would have to be made.

\subsection{Delimitation}
This paper is focused on revocation of VCs  in SSI systems. Revocation is also important in other areas as Internet of Things (IoT). Privacy- and technical requirements can be different in these areas. The focus is on cryptographic revocation methods and non-cryptographic revocation methods which can be used for revocation. Revocation is not combined with other cryptographic protocols like BBS group signatures \cite{boneh_short_2004} or primitives like Zero-Knowledge Proofs (ZKPs) even though they are necessary to create a full privacy preserving proof in an SSI system. 
\section{Contribution}
The paper provides the following contributions:
\begin{itemize}
\item It provides an overview over revocation methods and a classification in different revocation groups
\item A new revocation method called Linked Validity Verifiable Credentials (LVVC) is defined and described
\item Privacy- and technical requirements for a revocation method for VC in an SSI system based on a survey within the SSI practice are defined
\item An assessment framework for revocation methods is defined and can be used to assess other revocation methods
\item An assessment of the revocation methods to determine the applicability is provided
\item The shortcoming and open areas with unsolved problems are highlighted
\end{itemize}

\section{Prior Work and Grouping} \label{Prior Work and Grouping}
This chapter gives an overview about revocation methods and organize them in groups for the assessment.
\subsection{List Based}

List based revocation methods are the simplest method. There are different implementations that differ in their properties.\\
\\
\textbf{Simple List Based:} List based revocation methods are simple allow- or blocklists. The lists are publicly available or can be queried via an Interface without restrictions.  Examples are Certification Revocation Lists (CRL) in the X.509 certificate standard used for TSL \cite{noauthor_x509_nodate}. \\
\textbf{List Based Hidden:} In hidden list based revocation methods, the allow- or blocklist is hidden. A trusted party is necessary to manage the list and the access. Group Signatures can be used. A group manager controls the list and ensures anonymity\cite{chaum_group_1991}.\\
\textbf{Compressed List:} Compressed Lists compress the information. The advantage is that the size of the list is smaller, and the storage, download, and query can be done more efficiently. Bloom filters\cite{nyberg_fast_1996,acar_accumulators_nodate} or bit arrays can be used to implement compressed list methods.

\subsection {Cryptographic Accumulators}
An accumulator is a one-way function that sums a large set of items into one single accumulator value. The membership of an included item can be proofed with the accumulator, the item itself, and a witness file \cite{benaloh_one-way_1994,fazio_cryptographic_2002}.\\
\\
\textbf{Asymmetric and Symmetric accumulators:} Asymmetric accumulators need additional information, called a witness, for verification. Symmetric accumulators work without witness information. Examples for asymmetric accumulator are RSA based \cite{benaloh_one-way_1994, baric_collision-free_1997, camenisch_dynamic_2002,wang_new_2007,li_universal_2007, acar_accumulators_nodate}, Elliptic Curve/Bilinear Pairing based \cite{nguyen_accumulators_2005, camenisch_accumulator_2009,vitto_dynamic_2020} and Merkle tree based \cite{camacho_strong_2012, reyzin_efficient_2015, jhanwar_trading_2019} accumulators. An example for a Symmetric Accumulator is a Bloom Filter.\\
\\
\textbf{Update properties:} Cryptographic accumulators have different update properties. An update is an addition or/and deletion of an item. The definition is based on \cite{baldimtsi_accumulators_2017}.  A static accumulator cannot be updated. Therefore, a static accumulator is not suitable for revocation. An additive accumulator can only include additional items. Items that are included can never be removed from the accumulator. A subtractive accumulator can only exclude items. A dynamic accumulator is additive and subtractive. Items can be included and excluded. A dynamic accumulator allows excluding and thereafter include the same element.\\
\\
\textbf{Proofs:} Cryptographic accumulators have different proofing properties. The definition is based on \cite{baldimtsi_accumulators_2017}. A positive accumulator  can prove the membership of a certain item in an accumulator called membership proof. A negative accumulator can prove that an item is not included in an accumulator, called non-membership proof. A universal accumulator supports both membership and non-membership proofs. It must be possible for a verifier to verify the proof.
 \subsection{Credential Update}
The validity of a credential is time-limited. If the issuer chooses a short period of validity, e.g. 24h. The Holder must update his credential every 24h to keep validity. The Issuer issues the credential again with a new issuance date  \cite{camenisch_solving_nodate}.

\section{A new Revocation Method: Linked Validity Verifiable Credentials }
\begin{figure}[H]
    \centering
    \includegraphics[width=0.5\textwidth]{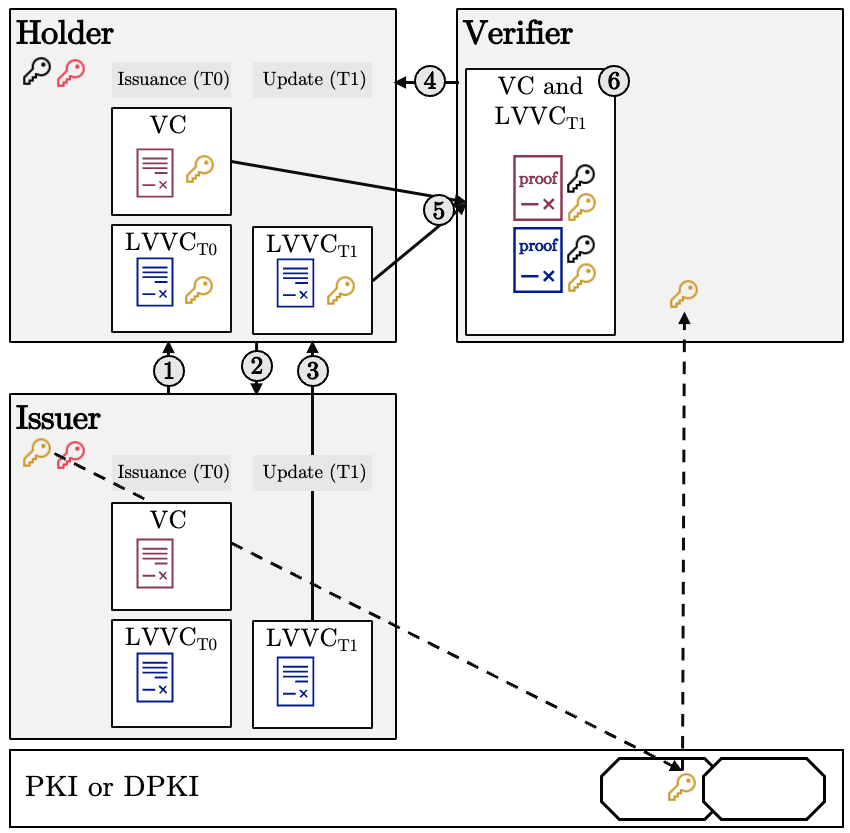}
    \caption{Process Flow LVVC}
    \label{fig:processFlowLVVC}
\end{figure}

Linked Validity Verifiable Credentials (LVVC) are introduced in this paper as a new revocation method.  A LVVC is a further development based on the principles of the credential update. A LVVC is a VC linked to another VC. The LVVC includes minimal information about the issuer and the time of issuance. The time of issuance is important for the determination of the validity. The Holder needs to update the LVVC on regular base depending on the requirements of the Verifier. The advantage compared to a re-issuance of the linked VC without an additional LVVC \cite{camenisch_solving_nodate} is the fixed size of the LVVC. A VC can vary in size and this can affect the scaleability. Another advantage of the decoupling is additional security and privacy. The update service from the Issuer does not need all information about the issued VC. Only the VC ID, Holder ID, Issuer signing key and the revocation status is needed to issue a new LVVC.\\
\\
In Figure \ref{fig:processFlowLVVC} the process flows of issuance, revocation and verification is described. The figure includes a layer which represents a PKI or a DPKI. The PKI or DPKI  acts as a trust layer where the Verifier can obtain information about the Issuer and the type of VC to be able to verify a VC and the LVVC.
\begin{itemize}
\item \textbf{Issuance} Step 1: an additional LVVC, together with the VC is issued to the Holder.
\item \textbf{Revocation/Update} Step 2,3: Revocation is done indirectly with the LVVC. The LVVC includes an issuance date. If the Holder needs a current date, he needs to request a new LVVC. If the VC is not revoked, the Issuer issues a new LVVC with the current issuance date.
\item \textbf{Verification} Step 4,5,6: The verifier initiates a verification request. The Holder creates a proof with the VC and LVVC. The Verifier performs the following verifications: Are the proofs validly signed? Is the VC and LVVC linked?  Does the issuance date of the LVVC meet the conditions?
\end{itemize}

\newpage
\section{Survey} \label{Survey}
A survey to define the baseline and minimal requirements for the following assessment was carried out. The survey consists out of two parts. The first addressed privacy (correlation, linkage and knowledge of transaction data) and the second part technical requirements (storage space, computational effort, network). We conducted the survey with members of the SSI practice, which are working on SSI implementations, are part of an SSI consortium and/or are involved in standardization activities. We invited an explanatory sample to the survey.  The survey was not anonymous to ensure the qualification of the participants. To be considered in the evaluation, the participates had to provide an email address, name and company/consortium. The information was checked, and if the participants could not be verified, the answers were not considered in the evaluation.
\subsection{Structure}
A semi-structured approach was applied. The composition of the questions consists of thirteen closed and required multiple-choice questions and eight open text fields where participants could add additional information.\\
\\
\textbf{Privacy:} The importance of data correlation, data linkage and traceability and the avoidance of collecting transaction data was queried. The participants have been asked if a violation of a privacy aspect from a revocation method would be a reason not to use a revocation method.\\
\\
\textbf{Technical}: The maximum acceptable storage space for three different roles in the revocation process, the Issuer, the Holder and the Verifier was asked because each role use different hardware. An Issuer will run the service most probably on a server, the Holder will manage the credentials on a mobile phone, and verification will be done via web services. Therefore, it is important to distinguish between the roles.
As well as the maximum acceptable storage space, the acceptable computational effort depending on the role was queried too. Time is more crucial during a verification process than for an update process in the backend of the Issuer. Also, the acceptable network bandwidth for an Issuer was included in the survey. Some revocation methods require transmission of additional information to each Holder. This information is often called witness. If the Issuer manages several millions of VCs in one accumulator, the amount of data can be significant, and therefore the network bandwidth has to be considered.
\newpage
\subsection{Results}
23 participants answered the survey.\\
\\
\textbf{Privacy:} The evaluation of the survey shows a uniform result in the privacy part. The questions in (figure 1) shows that for 78\% - 82\% privacy is important. 68\% would not use a revocation method which violate the three asked privacy aspects.
\begin{figure}[H]
    \centering
    \includegraphics[width=\textwidth]{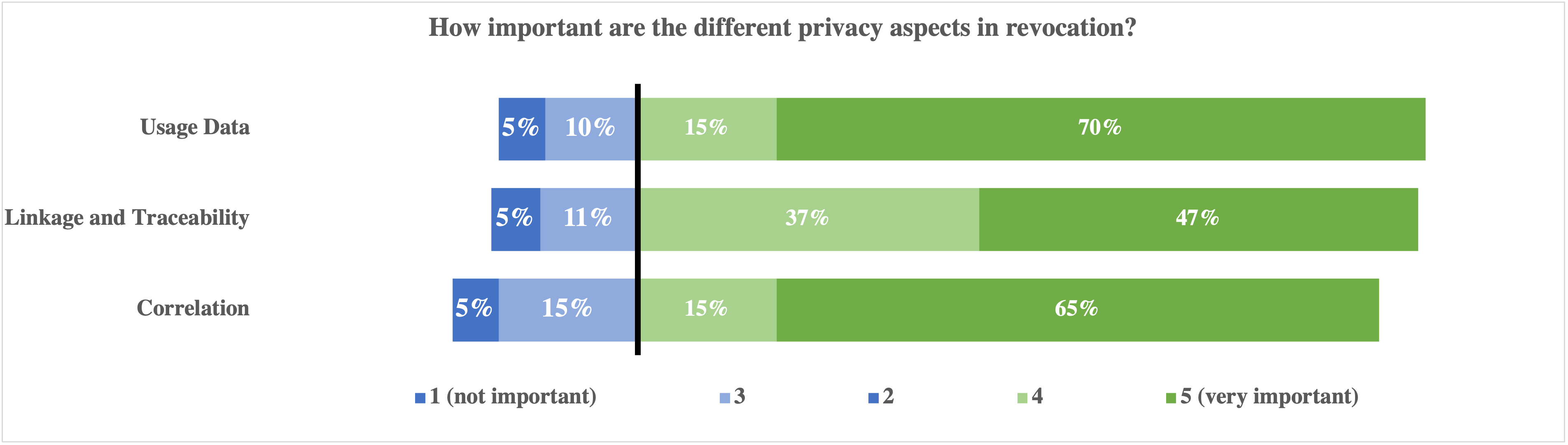}
    \caption{Privacy}
    \label{fig:privacyResults}
\end{figure}

\begin{figure}[H]
    \centering
   \includegraphics[width=\textwidth]{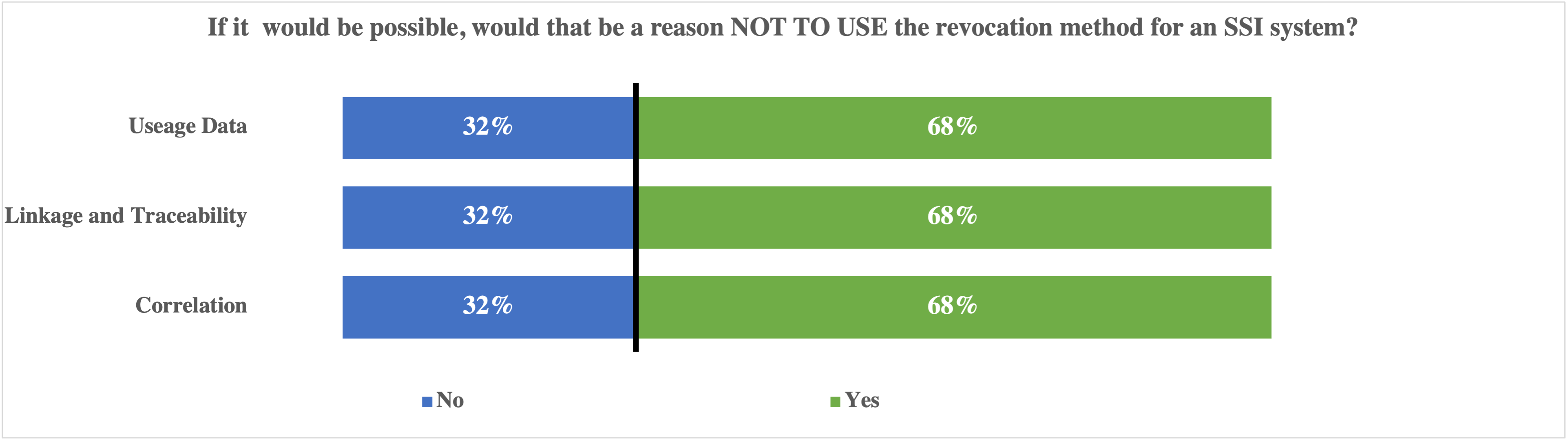}
    \caption{Privacy KO criteria}
    \label{fig:privacyKO}
\end{figure}
\newpage
\textbf{Technical:} The evaluation of the questions about the technical requirements does not show a consistent result. The answers are spread across the spectrum, with a high percentage of "Don't know" responses. We decide to use the Median as our baseline for our assessment.\\
\\
\textbf{Storage Space}
\begin{figure}[H]
    \centering
    \includegraphics[width=\textwidth]{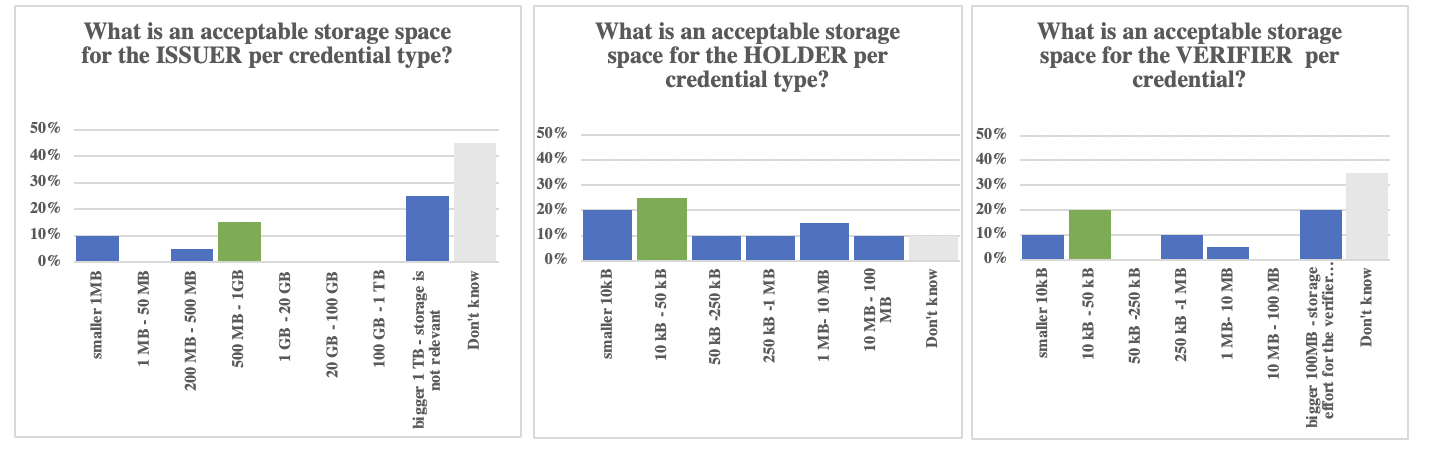}
    \caption{Requirements Storage Space}
    \label{fig:reqStorageSpace}
\end{figure}

\textbf{Computational Effort (time)}
\begin{figure}[H]
    \centering
    \includegraphics[width=\textwidth]{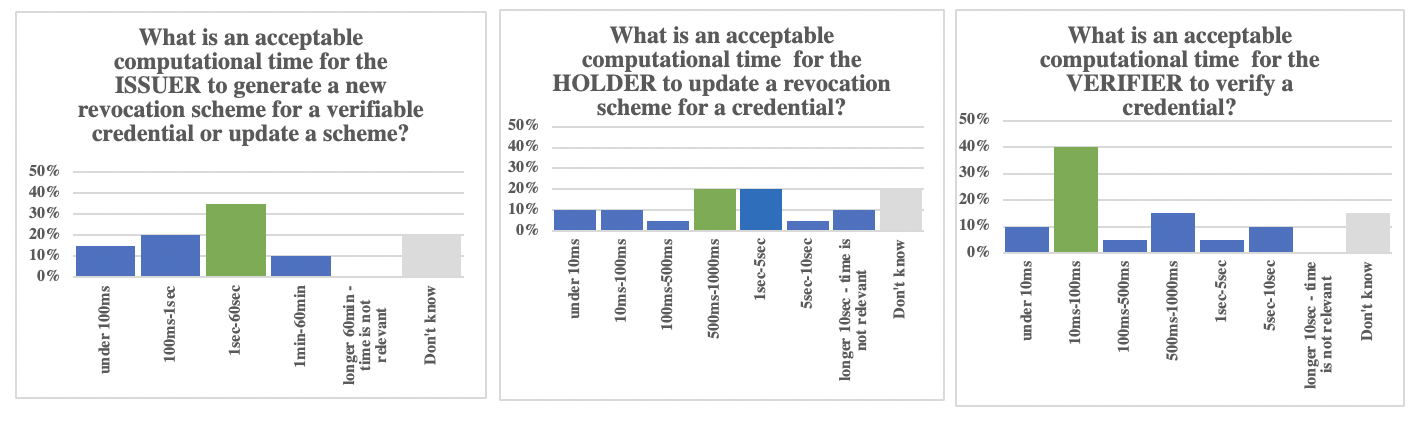}
    \caption{Requirements Computational Effort}
    \label{fig:reqComputationalEffort}
\end{figure}

\section{Requirements}
The requirements are not only based on technical parameters. An important aspect and pre-condition of SSI is the privacy of the Holder and Verifier. A revocation method must be privacy preserving to fulfil the principles of SSI. Therefore, the assessment is focused on privacy.\\
\subsection{Privacy}
To prove that privacy is a key characteristic in SSI systems, the history of SSI and the characteristic described has to be considered. In the year 2005 Cameron \cite{cameron_laws_2005} wrote a paper called "The Laws of Identity". The paper describes "laws" which must be fulfilled to raise the acceptance of DI systems. In central systems is that the user has to trust a central system. The compliance cannot be controlled from the outside. With reference to Cameron \cite{cameron_laws_2005} Christopher Allen wrote in 2016 a highly regarded blog post "The Path to Self-Sovereign Identity" \cite{allen_path_2016} where he described 10 principles of SSI as a starting point for further discussion. Satybaldy, Nowostawski and Ellingsen \cite{satybaldy_self-sovereign_2020} developed in the paper "Self-Sovereign Identity Systems: Evaluation Framework" an evaluation framework based on the work from Allen \cite{allen_path_2016}, Cameron \cite{cameron_laws_2005} and added usability as an additional requirement. They defined overall eight characteristics for SSI systems. 2020 Naik and Jenkins \cite{naik_governing_2020} developed in the paper "Governing Principles of Self-Sovereign Identity Applied to Blockchain Enabled Privacy Preserving Identity Management Systems" a more comprehensive evaluation framework with 20 characteristics. Naik and Jenkins applied it to the Sovrin SSI framework.
2019 Ferdous, Chowdhury and Alassafi \cite{ferdous_search_2019} introduced a formal model to describe and assess digital identity systems.  In the literature describing SSI, privacy is a key topic.\\
The results from the survey in chapter \ref{Survey} show that all three defined privacy aspects are relevant. For 78\% - 82\% privacy is important. 68\% would not use a revocation method which violate the three asked privacy aspects.

\subsection{Scalability and Maturity} \label{scalabilityAndMaturity}
An important requirement is scalability. A revocation method in an SSI system needs to be able to cover millions of credentials to compete with existing central based identity systems. In the paper, scalability is defined as usable in real world large scale ($>$ 1 million user and $>$1 million credentials) implementations. The results from the survey for computational effort and storage requirements in chapter \ref{Survey} are used as a base for the assessment.\\
In the assessment, the maturity is rated on base of existing implementations. If the method is used in large scale implementations then it is rated as high. The maturity of a method which is not in productive use is considered as low if needed primitives are at the research level and are not in a standardization process. If the underlying primitives are in a standardization process and are in use they but not in large scale implementations are rated as medium.
\clearpage
\section{Definition of Assessment Criteria} \label{Definition of Assessment Criteria}
In this chapter interactions and processes, privacy levels and privacy aspects which are relevant for the assessment are defined.
\subsection{Interactions and Processes} \label{Interactions and Processes}
Each revocation method requires interactions between Issuer, Holder, and Verifier.  An interaction is defined as a contact between different roles. Interactions in the issuance-, revocation- and verification process are assessed. The issuance process is the first issuance of a VC and additional steps necessary to enable revocation. In the revocation process, items will be revoked and data will be updated. In the verification process, the Verifier verifies the validity of an VC.
\subsection{Privacy Aspects} \label{Privacy Aspects}
The assessment includes three privacy aspects. Privacy aspects are assessed from the Holder perspective.
\begin{itemize}
\item \textbf{Correlation:} If one party is in possession of information, that gives the possibility to link the information with information from other parties. This can be anything which represents a unique identifier such as a specific HASH value, an identifiable credential or a public key that is used multiple times.
\item \textbf{Transaction data:} transaction data is metadata about the usage of a VC. 
\item \textbf{Linkage:} A revocation has to be proofed at a certain point in time. The Verifier should be able to verify the validity of the VC exactly at this point. Linkage is defined as the possibility from the Verifier to check the validity of the VC in the past or in the future without the Holder's involvement.
\end{itemize}

\subsection{Privacy Levels and Perspectives} \label{Privacy Levels and Perspective}
In an SSI system, the Holder interacts with two different parties, the Verifier, and the Issuer. Therefore, the privacy level is assessed from both perspectives. Privace levels are influenced from the interactions described in chapter \ref{Interactions and Processes}.\\
\\
\textbf{Privacy Level Holder to Issuer:}\\
Revocation methods have different privacy levels in the relationship between Holder and Issuer. 
\begin{itemize}
\item \textbf{Full Privacy:} The Issuer gets no information about the usage of the VC or/and the verifier.
\item \textbf{Semi Privacy:} The Issuer gets information that the VC is used from the Holder or/and gets the information that a Verifier is performing a validation process on a VC issued from the Issuer.
\item \textbf{No Privacy:} The Issuer knows the Holder and the Verifier in a validation process.
\end{itemize}
\newpage
\textbf{Privacy Level Holder to Verifier:}
Revocation methods have different privacy levels in the relationship between a Holder and a Verifier.
\begin{itemize}
\item \textbf{Full Privacy:} Full privacy is provided from anonymous revocation methods. Anonymous methods reveal only the current validity and nothing about the use or validity in the past and future.
\item \textbf{Semi Privacy:} Semi privacy methods do not reveal everything to the Verifier or public. An additional piece of information is needed for verification and access.
\item \textbf{No Privacy:} The validity of a credential can be verified by everybody, without restriction, and every time.
\end{itemize}

\section{Assessment}
The defined revocation groups in chapter \ref{Prior Work and Grouping} are assessed with the defined assessment criteria described in chapter \ref{Definition of Assessment Criteria}.

\subsection{Required interactions} \label{Required interactions}
The definition can be found in chapter \ref{Interactions and Processes}"Interactions and Processes". The analysis of the interactions is necessary for the determination of the privacy levels.\\
{
\rowcolors{2}{lightgray!30!}{lightgray!10!}
\begin{table}[H]
\begin{tabular}{ |p{5cm}|p{1cm}|p{1cm}|p{1cm}|p{1cm}|p{1cm}|p{1cm}|p{1cm}|p{1cm}|p{1cm}|}
\hline
&\multicolumn{3}{|c|}{Issuance} 
&\multicolumn{3}{|c|}{Revocation} 
&\multicolumn{3}{|c|}{Verification}\\
\hline
Group                           &I  &H  &V  &I  &H  &V  &I      &H  &V\\
\hline
List Based                      &x  &   &   &x  &   &   &x      &x   &x\\
List Based Hidden               &x  &   &   &x  &   &   &x    &x   &x\\
Compressed List                 &x  &   &   &x  &   &   &x      &x   &x\\
Cryptographic Accumulators      &x  &x  &   &x  &x  &   &       &x   &x\\
Credential Update               &x  &x  &   &x  &x  &   &       &x   &x\\
LVVC                            &x  &x  &   &x  &x  &   &       &x   &x\\
\hline
\end{tabular}
\caption{\label{tab:Interactions}Required Interactions}
\end{table}
}
Table \ref{tab:Interactions} shows, that list based revocation methods have an advantage in the issuance and revocation process. No interaction between any other role is required because of revocation. But there are disadvantages in the verification process. In a list based approach, the Issuer or another third party needs to be contacted to validate the VC, this is called "calling-home".\\
In non-list based revocation methods, the Verifier does not contact the Issuer during the verification process. To make this possible, the Holder requires additional information from the Issuer, the witness. The witness is issued to the Holder during the initial issuance process and needs to be re-issued or updated after each revocation process.  Cryptographic accumulator methods needs to update all witness information from all contained VCs. With the credential update or LVVC method, not all witnesses are effected. An update is only performed on VCs where the revocation status is changed.
\subsection{Holder Privacy}
Based on the interactions defined in chapter \ref{Required interactions}, the privacy aspects from the Holder towards the Issuer and towards the Verifier are assessed. For the Holder to Verifier perspective, only correlation and linkage are relevant, as the Verifier is never involved when another Verifier verifies a VC. For the Holder to Issuer perspective, only the transaction data aspect is relevant. Linkage is not relevant for the Issuer as he knows all revocation information from his issued VCs and if the Issuer has transaction data he does not need to correlate the data. The assessment of the privacy aspects define the privacy level. Full privacy is only possible if correlation, linkage is not possible and transaction data cannot be collected. In the assessment, we assume a correct implementation of the method. Every privacy preserving method can be implemented less privacy preserving. Cells marked with "y-n" reflect the dependency on a privacy preserving implementation, e.g.use of zero knowledge proofs and no-unique identifiers.\\
Cryptographic Accumulators, credential updates and LVVC can provide full privacy as the Issuer learns nothing about the usage of the VC.\\
Table \ref{tab:Holder Privacy} shows for Holder to Issuer privacy, that all list based methods provide No Privacy. Every list based method require the Verifier to contact the Issuer and therefore the Issuer knows when and from which Verifier the VC is used. 
Holder to Verifier privacy depends also on the design of the system. If the system is using unique identifiers or generated proofs are always the same, it is possible for different Verifiers to correlate the data.\\
Linkage is possible with list based methods, but not with the non list based revocation groups. The data provided for validation is changing if the revocation status of one or more items changes. Therefore, the revocation status for an VC can only be checked for one point in time and not before this point or in the future.
{
\rowcolors{2}{lightgray!30!}{lightgray!10!}
\begin{table}[H]
\begin{tabular}{ |p{5cm}|p{2cm}|p{2cm}|p{2cm}|p{2cm}|p{2cm}|}
\hline
&\multicolumn{2}{|c|}{Holder to Issuer}
&\multicolumn{3}{|c|}{Holder to Verifier}\\
\hline
Group   &Transaction data &Privacy Level	&Correlation  &Linkage &Privacy Level \\
\hline
List Based                    				   &y  &No Privacy	&y      	&y   &No Privacy     \\
List Based Hidden              		   &y  &No Privacy	&y-n	  &y-n &Semi Privacy  \\
Compressed List                		   &y  &No Privacy	&y      	&y   &Semi Privacy   \\
Cryptographic Accumulators	 &n  &Full Privacy	&y-n    &n   &Full Privacy  \\
Credential Update              		   &n  &Full Privacy	&y-n    &n   &Full Privacy  \\
LVVC                           				    &n  &Full Privacy	&y-n    &n   &Full Privacy  \\
\hline
\end{tabular}
\caption{\label{tab:Holder Privacy}Holder Privacy}
\end{table}
}

\subsection{Scalability and Maturity}
The basis for the assessment is defined in chapter\ref{scalabilityAndMaturity}.\\

\textbf{List Based Methods:} List based methods and List Based Hidden methods are used in X 5.09 certificates with Certification Revocation List \cite{noauthor_x509_nodate} or Online Certificate Status Protocol (OCSP)\cite{noauthor_x509_nodate}. Therefore, the scalability and maturity is defined as high in the table above.\\
\\
\textbf{Compressed List:} The scalability and maturity of Compressed List methods depends on the implementation. If Merkle Trees and bit-arrays are used, the scalability and maturity is high, as these are used in production systems.\\
\\
\textbf{Cryptographic Accumulators:} The only cryptographic accumulator method currently in use for SSI systems is the RSA based accumulator in the Hyperledger Indy project \cite{noauthor_0011_nodate}. A pilot implementation in the Province of British Columbia in Canada has revealed limitations  regarding witness file size and computational effort\cite{noauthor_indy_2022}. 32.768 VC included in the accumulator lead to a tail-file size of 8,4 MB and a proof generating time of 7 sec. with an iPhone12. The file needs to be downloaded from the Holder and is needed to calculate the witness. The file size grows linear with the number of VCs included. This proofs that this specific method does not scale in a large scale environment. Cryptographic Accumulators are not used in large scale production environments and have limitations in scalability. Therefore this method is rated low regarding scalability and mid-low regarding maturity. The working group applied crypto in the Decentralized Identity Foundation (DIF) is working on new cryptographic accumulator methods \cite{noauthor_dif_2022} with the goal to improve scalability. The work is at the beginning and additional work in research and implementation needs to be done.\\
\\
\textbf{Credential Update and LVVC:} For a VC, JavaScript Object Notation (JSON) \cite{bray_javascript_2014} as format standard and JSON Web Token\cite{jones_json_2015} for signature can be used. These are used and established standards. Therefore, the maturity of revocation with a Credential Update and LVVC is rated high. The scalability depends on the size of the VC. In the Credential Update method, the size of the VC cannot be controlled. The Issuer can integrate unlimited attributes and data, therefore the scalability is rated as high-medium. The LVVC always has the same size. A JSON VC signed with BBS \cite{boneh_short_2004} based is $<$1100 bytes. If an Issuer needs to manage 1 million credentials, 1,1 Gigabyte of storage are needed. The payload transmitted to the Holder during an update is only $<$1100 bytes. Therefore, the scalability for the LVVC method is rated high.
{
\rowcolors{2}{lightgray!30!}{lightgray!10!}
\begin{table}[H]
\begin{tabular}{ | p{5cm} | m{5cm} | m{5cm} |}
\hline
Group                          						 &Scalability  &Maturity\\
\hline
List Based                    					   &high        &high\\
List Based Hidden              			   &high        &high\\
Compressed List                				&high        &high\\
Cryptographic Accumulators       &low        &medium- low\\
Credential Update               		  &high-medium &medium\\
LVVC                          					   &high        &medium\\
\hline
\end{tabular}
\caption{\label{tab:scalability and maturity}Scalability and Maturity}
\end{table}
}

\section{Discussion and Conclusion}
The perfect revocation method would be a method where interactions are reduced to a minimum.  During issuance, no additional interactions should be necessary, as with list based revocation methods. During a revocation update, no interaction between the Issuer and the Holder should be necessary, as with list based methods and during verification there should be no interaction towards the Issuer, as with non-list based methods. At the time being, no revocation method combines all these interaction requirements.\\
The best method for privacy reserving revocation are cryptographic accumulators combined with zero knowledge proofs. The assessment showed that cryptographic accumulators have limiters in scalability and maturity. More work needs to be done in  research, implementation and testing.
Therefore, the LVVC method is the most suitable for SSI implementations. LVVC offer better privacy than list based methods and can provide the needed scalability and maturity. 
\newpage
\printbibliography
\end{document}